\documentclass[prd,floatfix,preprintnumbers]{revtex4}
\usepackage{amsmath}
\usepackage{graphicx}% Include figure files
\usepackage{dcolumn}% Align table columns on decimal point
\usepackage{newtxtext,newtxmath}
\usepackage[lofdepth,lotdepth]{subfig}
\usepackage{float}% If comment this, figure moves to Page 2

\newcommand {\ga} {\ {\raise-.5ex\hbox{$\buildrel>\over\sim$}}\ }
\newcommand {\la} {\ {\raise-.5ex\hbox{$\buildrel<\over\sim$}}\ }
\newcommand {\bfx} {\bf x}

\def\be{\begin{equation}}
	\def\ee{\end{equation}}
\def\ba{\begin{eqnarray}}
	\def\ea{\end{eqnarray}}
\def\hata{\hat{\bf a}}
\def\hatb{\hat {\bf b}}
\def\Ci{{\rm Ci}}

\begin{document}
	\title{Gravitational Radiation Power Spectrum of Garfinkle-Vachaspati Cosmic String Loops}
	\author{S. David Storm and Robert J. Scherrer}
	\affiliation{Department of Physics and Astronomy, Vanderbilt University,
		Nashville, TN  ~~37235}
	
	\begin{abstract}
		We examine the power spectrum $P_n$ of the $n$th harmonic
		for Garfinkle-Vachaspati cosmic strings, which correspond to planar rectangular
		loops.  While these loop tractories are self-intersecting, a slightly perturbed non-self-intersecting
		form of these trajectories
		has been suggested as a generic end state of cosmic string loop fragmentation.  We find that
		$P_n$ scales as $n^{-2} \ln n$, rather than the expected $n^{-2}$ for kink-kink collisions.  This
		result is demonstrated analytically for even $n$ in square loops and numerically for all other cases.
		At lowest order, the effect of loop decays is to further enhance $P_n$ at large $n$ relative to its value
		at small $n$.
		The consequences for relic stochastic background radiation along with the caveats pertaining to our results
		are discussed.
	\end{abstract}
	
	\maketitle
	
	\section{Introduction}
Cosmic strings are hypothetical one-dimensional topological defects that might
have formed at a cosmological phase transition in the early universe (see,
e.g., Ref. \cite{hep-ph/9411342} for a review).  As they
serve as copious emitters of gravitational radiation, there has been a recent revival of interest
in these objects.  Roughly speaking, there are three plausible signatures of this radiation.  First,
cusps on the strings (moving near the speed of light) can emit strongly beamed gravitational radiation.
Second, the gravitational radiation from the entire collection of cosmic strings can produce a stochastic background.  (See,
e.g., Refs. \cite{gr-qc/0104026,Blanco-Pillado} for a discussion of these two mechanisms).
Third, it has been argued that the pattern of radiation from a single oscillating cosmic string loop is
sufficiently distinctive that it might be observable in future detectors \cite{0904.1052,2006.00438}.  This last possibility has received considerably less attention than
the other two, although there have been limited attempts to detect such a signal \cite{2010.06118}.
More recently, a hybrid of the first and third signatures has been proposed:  a single cosmic string could
produce repeated gravitational wave bursts from cusps \cite{Auclair}.  In this paper, we will be concerned
with the stochastic background and the power spectrum of radiation that produces it.

It is thought that the initial configuration of cosmic strings produced in a phase transition
would consist
primarily of infinitely-long strings, with a smaller fraction in the form of
closed loops \cite{TUTP-84-1}. Subsequent evolution leads to a characteristic distribution of cosmic string loops as the loops are chopped off of the infinite strings and subsequently fragment.  It is these loops which are the dominant source of gravitational radiation from cosmic strings.  Detailed simulations of loop
fragmentation were performed by Scherrer and Press \cite{SP}, Copi and Vachaspati \cite{CV}, and Blanco-Pillado,
Olum, and Shlaer \cite{BOS1}.  The results of Ref. \cite{CV}, in particular,
suggest that the generic end state of such fragmentation is a set of nearly planar quasi-rectangular
loops, closely resembling the class of degenerate loop trajectories first explored by
Garfinkle and Vachaspati \cite{1987PhRvD..36.2229G}, which we shall call ``GV loops."
A somewhat different set of loops were obtained in Ref. \cite{BOS1}; we discuss these differences in Sec. V below.
However, in this paper, we explore the possibility that the generic end state of cosmic string
loop fragmentation is something resembling a collection of GV loops.  Of course, physical cosmic string loops cannot have exactly the GV form, since such
trajectories are highly degenerate and therefore self-intersecting, but it was pointed out in Ref. \cite{CV} that
slightly perturbed GV loop trajectories can be stable against self-intersection. 
Here we take the GV loop as a proxy for the generic final state of fragmentation and explore the consequences of
this assumption
for gravitational wave signatures.  While the total power emitted per unit solid angle of the gravitational radiation for GV loops
was investigated in the original paper by Garfinkle and Vachaspati \cite{1987PhRvD..36.2229G}, the quantity relevant for the stochastic
background is the total power $P_n$ as a function of the $n$th harmonic.  In this paper, we calculate $P_n$ as a function of $n$
for the GV loops, and then examine the lowest order effects of
 gravitational back-reaction.  Even
if the GV loops turn out to represent a poor approximation to the final loop distribution, their simplicity allows
for the derivation of some exact results that may provide insight into the behavior of the true final loop
distribution.

In the next section we summarize the evolution of cosmic strings loops oscillating well inside the horizon and review the
details of the GV loops.  In Sec. III, we calculate the initial power spectrum $P_n$ for the GV loops both analytically
(for square loops and even $n$) and numerically (for all others).  In Sec. IV, we examine how loop decay alters the form for $P_n$.  The
implications of our results are discussed in Sec. V.

\section{Production of GV Loops in Cosmic String Evolution}

The formation and evolution of cosmic strings has long been a topic of interest.  Early simulations \cite{TUTP-84-1} suggested that 
a phase transition would produce a network of strings consisting mostly of infinite strings, with a smaller proportion of closed loops.
In the standard picture of cosmic string evolution, loops continuously break off from the infinite strings and then fragment into smaller
loops, so that the network at any given time is a mixture of infinite strings and closed loops.
The closed loops decay via gravitational radiation and are the dominant source of this radiation in the cosmic string
scenario.

The actual evolution of the full network of cosmic strings has been investigated in detail with increasingly sophisticated analytic calculations
and numerical simulations \cite{AT1,Bennett1,Bennett2,BB1,BB2,AT2,CA1,Siemens1,Siemens2,VOV1,Martins,VOV2,Polchinski1,Polchinski2,Ringeval,OV,Copeland,Lorenz,BOS2}.
Despite all of this work, uncertainties remain regarding the exact evolution of the string network.  Here, however, we will be interested
in a simpler question:  the evolution of string loops well inside the horizon.

The cosmic string equation of motion for a closed loop well inside the horizon is
\begin{equation}
		\ddot x^\mu - x''^\mu=0,
		\label{eom}
	\end{equation} 

with

	\begin{equation}
		\dot x^\mu x'_\mu = 0,
	\end{equation}
and
	\begin{equation}	
		 \dot x^\mu \dot x_\mu +x'^\mu x'_\mu = 0. \label{eq:constraints}
	\end{equation} 
Taking $x^0 = \tau$, the solution for the three-vector $\bfx$ is
	
	\begin{equation}
		{\bf x}(\sigma,t)=\frac{1}{2}\Big[{\bf A}(t+\sigma)+{\bf B}(t-\sigma)\Big],
		\label{eomSolution}
	\end{equation} 
where ${\bf A}$ and ${\bf B}$ satisfy
${\bf A}^{\prime 2} = {\bf B}^{\prime 2} = 1$ and ${\bf A}(\sigma + L) = {\bf A}(\sigma)$, ${\bf B}(\sigma + L) = {\bf B}(\sigma)$,
with $L$ being the length of the loop.  These trajectories can contain points at which the speed of a point on the loops is
instantaneously equal to the speed of light, corresponding to ${\bf A}^\prime(u) = {\bf B}^\prime(u)$ for some value of $\tau$.  These points
are called cusps.  Alternatively, the string trajectory can contain discontinuities in the values of $A^\prime$ or $B^\prime$.  These
arise generically as a result of loop fragmentation and are called kinks.

A number of authors have explored the evolution of string loops well inside the horizon, for which Eq. (\ref{eomSolution}) gives
the exact equation of motion \cite{SP,CV,BOS1}.  When these closed loops self-intersect, they generically fragment into two smaller loops, with a kink forming on both loops
at the instant of fragmentation. However, each kink evolves into a left-moving and a right-moving kink corresponding to a discontinuity in both $A^\prime$ and $B^\prime$.  
It is easy to show that multiple loop fragmentations yield an average number of kinks per loop that asymptotically approaches 4.
As each loop fragments, it produces two daughter loops, with a single kink on each loop at the instant of fragmentation.
Thus,
a single loop fragmentation produces one additional loop and 4 kinks, and $N$ fragmentations correspond to $4N$ kinks and $N+1$ daughter loops, so
that the average number of kinks per loop is $4N/(N+1)$, which asymptotes to 4 in the limit of large $N$ \cite{SQSP}.

\begin{figure}[!htbp]
		\centering
		\includegraphics[width=\textwidth, trim = 2cm 10cm 2cm 10cm, clip]{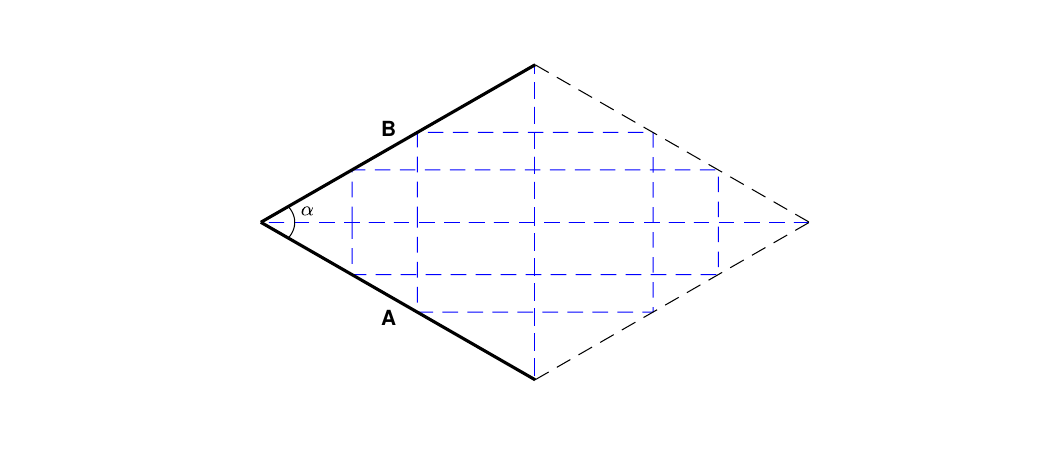}
		\caption{Evolution of a GV loop with the indicated opening angle $\alpha$.  At any given time, the loop
		corresponds to one of the blue dashed rectangles \cite{1609.01685}.}
		\label{fig:wh}
	\end{figure}

Furthermore, the results of Refs. \cite{CV} and \cite{BOS1} suggest that the final loops produced by fragmentation are highly planar.  Hence, Ref.
\cite{CV} suggested that the final state of the fragmenting loops could be well-approximated by quasi-planar, quasi-rectangular loops.  Exactly
planar, exactly rectangular loops were first explored by Garfinkle and Vachaspati \cite{1987PhRvD..36.2229G}.  For these
GV loops, both ${\bf A}$ and ${\bf B}$ are
closed line segments:
\begin{eqnarray}
{\bf A^\prime}(u) &=& \hat {\bf a}, ~~~~~~0 \le u \le L/2, \\
{\bf A^\prime}(u) &=& -\hat {\bf a}, ~~~L/2 \le u \le L,\\
{\bf B^\prime}(v) &=& \hat {\bf b}, ~~~~~~0 \le v \le L/2,\\
{\bf B^\prime}(v) &=& -\hat {\bf b}, ~~~L/2 \le v \le L,
\end{eqnarray}
where $\hat {\bf a}$ and $\hat {\bf b}$ are unit vectors, with
\begin{equation}
\hat {\bf a} \cdot \hat {\bf b} = \cos \alpha.
\end{equation}
Here $\alpha$ is the angle between the ${\bf A}$ and ${\bf B}$ line segments.  As noted in Ref. \cite{1609.01685}, there is an easy way to represent
the corresponding loop trajectories.  If a rhombus is constructed with opening angle $\alpha$, the string loop oscillates through all
rectangular configurations inscribed within the rhombus (see Fig. \ref{fig:wh}).
Although $\alpha$ can vary from $0$ to $\pi$, it is clear from Fig. \ref{fig:wh} that loop trajectories with $\pi/2 < \alpha < \pi$ map exactly
onto trajectories with $0 < \alpha < \pi/2$.  Hence, it will be sufficient to consider only $\alpha$ between $0$ and $\pi/2$.
(Note that we use the definition of $\alpha$ from Ref. \cite{1987PhRvD..36.2229G}, which differs from the opening angle
defined in Ref. \cite{1609.01685} by a factor of 2).

Of course, these GV exact trajectories cannot correspond to real physical loops, since they all collapse to a completely self-intersecting line during their
oscillation.  However, Copi and Vachaspati have constructed perturbed GV loops that are stable against self-intersection \cite{CV}.  Since
these quasi-GV trajectories can be made arbitrarily close to exact GV trajectories, it is not unreasonable
to explore the exact GV loops as proxies for the final distribution of loops in the simulation of Ref. \cite{CV}.

\section{Initial Power Spectrum of Gravitational Radiation from GV Loops}

The power spectrum of gravitational radiation from cosmic strings is of central importance in the calculation
of the stochastic background produced in such models.  In general, the density per logarithmic frequency
interval relative to the critical density $\rho_c$ is given by \cite{Blanco-Pillado21}
\begin{equation}
\Omega_{GW}(\ln f) = \frac{1}{\rho_c} G\mu^2 \sum_{n=1}^\infty P_n \frac{2n}{f} \int_0^{t_0} \frac{dt}{(1+z)^5}
N(L,t),
\end{equation}
where the integral is taken over the time up the present, and $N(L,t)$ is the number density
of loops of length $L$ at time $t$.  This integral depends on the details of the string network evolution, which
can be determined from numerical simulations, although it remains somewhat uncertain.  Here, however,
we are interested in the power spectrum $P_n$ of gravitational radiation produced by a given loop.

As noted by Damour and Vilenkin \cite{gr-qc/0104026}, cosmic string loops with cusps generically produce a power spectrum of the form $P_n \propto n^{-4/3}$ (see also
Ref. \cite{Binetruy} and earlier calculations in Ref. \cite{VandV}).
In the absence of cusps, loops with kinks give $P_n \propto n^{-5/3}$ \cite{1004.0890}, and loops with kink-kink collisions give $P_n \propto n^{-2}$ \cite{gr-qc/0104026,1709.03845}.  Thus, for large $n$, it is often assumed that loops with cusps will
dominate $P_n$, giving a contribution that scales as $n^{-4/3}$, so
it is this power spectrum that is generally used in calculations of the stochastic background predicted from cosmic strings.  However, Ringeval and Suyama \cite{1709.03845} and Olmez et al. \cite{1004.0890} have
challenged this assertion, with the former investigating the possibility that kink-kink collisions might dominate the resulting spectrum,
while the latter suggested that cusps and kinks might contribute equally.
Hence, when calculating
the stochastic background of gravitational radiation from cosmic strings, the actual form for $P_n$ is
sometimes taken to be a free power law or a sum of power laws,
 with the results dependent on this choice \cite{Blanco-Pillado21,nanograv,Kume,Schmitz}.

For the special case of GV loops,
the power in gravitational radiation was first calculated in Ref. \cite{1987PhRvD..36.2229G}.  The power emitted
at frequency $\omega_n = 4\pi n/L$ in the direction of the unit vector $\hat r$ is given by \cite{1987PhRvD..36.2229G}
\begin{equation}
\label{Pn}
\frac{dP_n}{d\Omega} = \frac{32 G\mu^2}{\pi^3 n^2} \frac{[1-(-1)^n \cos(n \pi e_1)][1-(-1)^n \cos(n \pi
e_2)]}{(1-e_1^2)(1-e_2^2)}, 
\end{equation}
where $e_1 = \hat r \cdot \hat{\bf a}$ and $e_2 = \hat r \cdot \hat{\bf b}$.
While Ref. \cite{1987PhRvD..36.2229G} was primarily concerned with the total integrated power, here we are interested
in the behavior of $P_n$ as a function of $n$.
We have numerically integrated Eq. (\ref{Pn}) over the sphere to derive $P_n$, which
is shown in Fig. 2 for a variety of values of $\alpha$.
	\begin{figure}[!htbp]
		\centering
		\includegraphics[width=\textwidth, trim = 2cm 8cm 2cm 8cm, clip]{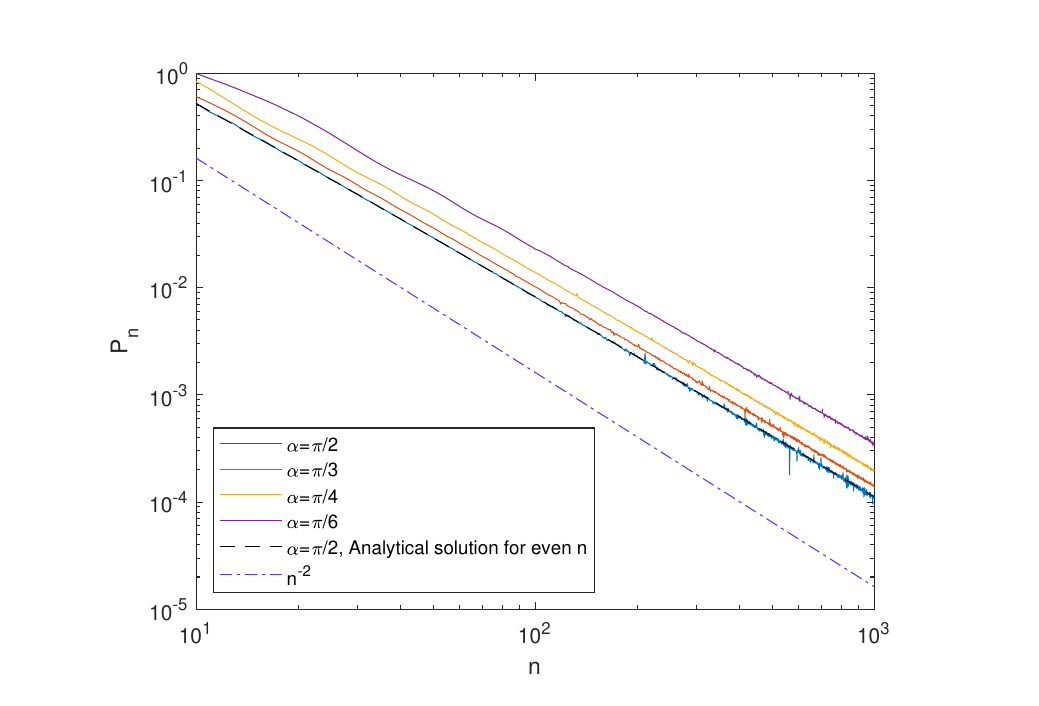}
		\caption{The power spectrum $P_n$ as a function of harmonic $n$ for Garfinkle-Vachaspati
		loops with the indicated opening angle $\alpha$.  Dashed curve is the analytic result for
		$\alpha = \pi/2$ in the large-$n$ limit (Eq. \ref{exactsol}).}
		\label{fig:worldsheet}
	\end{figure}
For the case of GV loops, we would expect $P_n$ to scale as $n^{-2}$, which is also displayed
in Fig. 2.  It is clear that $P_n$ does not scale as a pure $n^{-2}$ power law for the GV loops.

While an analytic derivation of $P_n$ is rather difficult in general, it is possible to integrate Eq. (\ref{Pn})
exactly for the
case of a square loop ($\alpha = \pi/2$) with $n$ even (see Appendix A).
The result, in the limit of large $n$, is
\begin{equation}
\label{exactsol}
P_n = \left[ \frac{128 G\mu^2}{\pi^2 n^2}\right ] \left[\gamma + \ln \pi
+ \ln(n)\right],
\end{equation}
where $\gamma$ is the Euler-Mascheroni constant $(= 0.5772)$.  This expression for $P_n$ is plotted in Fig. 2.
It is clear that Eq. (\ref{exactsol}) provides an excellent fit to the numerical results for $\alpha = \pi/2$.
The logarithmic term in Eq. (\ref{exactsol}) dominates the constant term for $n \ge 6$, so
the asymptotic behavior of $P_n$ in this case is $P_n \sim n^{-2} \ln(n)$, rather than the expected
$P_n \sim n^{-2}$.
While the loops with $\alpha < \pi/2$ are less amenable to analytic treatment,
it is obvious from Fig. 2 that $P_n$ for these loops also scales as $n^{-2} \ln(n)$, albeit
with a different multiplicative constant.  This appears to be the generic result for $P_n$ for GV loops.

	\section{The Power Spectrum from Decaying Loops}
The results of the previous section apply only to the initial state of the GV loops.  As is well known,
the emission of gravitational radiation produces a back reaction that alters the trajectory of the loops
and thereby changes the pattern of emitted power \cite{Quashnock1,Quashnock2}.  Analytic expressions for the
effect of the backreaction on piecewise linear loops have been worked out by
Wachter and Olum \cite{1609.01685} and applied in that paper specifically to the GV loops; we will make
use of their results here.
	
The GV loop is composed of four straight line segments.   In Ref. \cite{1609.01685},
it was noted that
each loop segment carves out a diamond-shaped worldsheet, and the gravitational backreaction was calculated at each point
on these ``diamonds.". Following this treatment, we will use $\tau$ and $\sigma$ for the temporal and spatial dimensions of the loop, respectively, $\tau_\pm=\tau\pm\sigma$, and move to diamond-specific coordinates $u= \tau_+-\tau_+^{(0)}$ and $v = \tau_--\tau_-^{(0)}$, where $\tau_+^{(0)}$ and $\tau_-^{(0)}$ are the values of $\tau_\pm$ at the center of the given diamond. For the square GV loop we will take the center of the diamond to occur at $\tau_\pm^{(0)}=\frac{L}{4}$ for $0<\tau_\pm<\frac{L}{2}$ and at $\tau_\pm^{(0)}=\frac{3L}{4}$ for $\frac{L}{2}<\tau_\pm< L$.
	
Taking $w = \cos(\alpha/2)$ and $h = \sin (\alpha/2)$, and defining $R=1+2(w/h)^2$ and $R'=1+2(h/w)^2$, the changes to the null tangent vectors $A'$ and $B'$ are given, for the diamond moving in the $+x$ direction (when $\tau_\pm<\frac{L}{2}$), by \cite{1609.01685}
\begin{eqnarray}
\Delta B'^\lambda &=& 8G\mu\Biggr( \log\Big[ \frac{(1-u)^2}{(R+u)(R'+u)} \Big],\frac{1}{w}\log\Big[\frac{1-u}{R+u}\Big],\frac{1}{h}\log\Big[\frac{1-u}{R'+u}\Big],0 \Biggr) \text{ and} \nonumber\\
\Delta A'^\lambda &=& 8G\mu\Biggr( \log\Big[ \frac{(1-v)^2}{(R+v)(R'+v)} \Big],\frac{1}{w}\log\Big[\frac{1-v}{R+v}\Big],-\frac{1}{h}\log\Big[\frac{1-v}{R'+v}\Big],0 \Biggr), 
\label{modifications}
\end{eqnarray} 
and the modifications for the remaining three diamonds can be found by multiplying $\Delta B'^\lambda$ or $\Delta A'^\lambda$ by $-1$ when $\frac{L}{2}<\tau_+$ or $\frac{L}{2}<\tau_-$, respectively. This rule for finding the remaining diamonds applies for the remainder of this section, and we will choose $L=4$ for the remainder of the analysis and calculations for simplicity.
	
	 As in \cite{1609.01685}, we take logarithmic divergences for $1-u$ or $1-v$ to be at least $\delta$ numerically, which corresponds to having the string pass by itself at a distance of $\delta$ in the $z$-direction during oscillation instead of proceeding with its double line formation. We will also make use of a change of coordinates from $u$ and $v$ to $\tilde{u}$ and $\tilde{v}$ such that the temporal modification becomes $0$ for each $\Delta B'$ and $\Delta A'$ \cite{1609.01685}. This new change of coordinates is found by calculating the fraction $(B'^\lambda+\Delta B'^\lambda)/(B'^0+\Delta B'^0)$ and using the binomial approximation $(1+x)^{-1} \approx 1-x$ when $x\ll 1$, and neglecting terms of order $(G\mu)^2$ squared \cite{thesis}. The explicit requirement is
	 
	 \begin{equation}
	 	\left|\Delta B'^0\right| = \left|NG\mu\log\Big[ \frac{(1-u)^2}{(R+u)(R'+u)} \Big]\right| \ll 1.
	 	\label{NGmurequirement}
	 \end{equation}
	 
	Then we have \cite{thesis}
\begin{eqnarray}
\Delta B'^\lambda &=& 8G\mu\Biggr( 0,\frac{1}{w}f(\tilde{u}),-\frac{1}{h}f(\tilde{u}),0 \Biggr), \nonumber\\
\Delta A'^\lambda &=& 8G\mu\Biggr( 0,\frac{1}{w}f(\tilde{v}),\frac{1}{h}f(\tilde{v}),0 \Biggr),
\label{reparam}
\end{eqnarray}
where 
\begin{equation}
f(n) = h^2\log\Big(\frac{1-n}{R+n}\Big)-w^2\log\Big(\frac{1-n}{R'+n}\Big),
\label{fundF}
\end{equation}
and
\begin{equation}
\tilde{u},\tilde{v} \in \Biggr(-1-8G\mu\Bigr[\frac{\log w^2 }{h^2}+\frac{\log h^2 }{w^2}\Bigr],1+8G\mu\Bigr[\frac{\log w^2}{h^2}+\frac{\log h^2}{w^2}\Bigr]\Biggr).
\label{scaling}
\end{equation} 

To determine $\Delta B^\lambda$ and $\Delta A^\lambda$, the integral of Eq. (\ref{fundF}) is found:
\begin{equation}
F(n) = w^2\Biggr[ (1-n)\log\Big(\frac{1-n}{R'+n}\Big) + (R'+n)\log(R'+n) \Biggr]
	- h^2\Biggr[ (1-n)\log\Big(\frac{1-n}{R+n}\Big) + (R+n)\log(R+n) \Biggr],
	\label{fundInt}
\end{equation} 
and the full equations for
$\Delta B^\lambda$ and $\Delta A^\lambda$
which include the constants of integration as $C^\lambda_B$ and $C^\lambda_A$, are
\begin{eqnarray}
\Delta B^\lambda &=& 8G\mu\Biggr( 0,\frac{1}{w}F(\tilde{u}),-\frac{1}{h}F(\tilde{u}),0 \Biggr) + C^\lambda_B \text{ and}  \nonumber\\
\Delta A^\lambda &=& 8G\mu\Biggr( 0,\frac{1}{w}F(\tilde{v}),\frac{1}{h}F(\tilde{v}),0 \Biggr) + C^\lambda_A.
	\label{DeltaAB}
\end{eqnarray}
The constants of integration are found by setting $\Delta B^\lambda(u = -\frac{L}{4})=\Delta A^\lambda(v = -\frac{L}{4})=0$ for $\tau_\pm<\frac{L}{2}$ as well as $\Delta B^\lambda(u = \frac{L}{4})=\Delta A^\lambda(v = \frac{L}{4})=0$ for $\frac{L}{2}>\tau_\pm$. This will ensure that the $B$ and $A$ loops are properly closed and centered. The result, again for the $+x$ diamond, is
\begin{eqnarray}	
C^\lambda_B &=& -8G\mu\Biggr(0,\frac{1}{w}F\Big(u = -\frac{L}{4}\Big),-\frac{1}{h}F\Big(u = -\frac{L}{4}\Big),0\Biggr),\\
C^\lambda_A &=& -8G\mu\Biggr(0,\frac{1}{w}F\Big(u = -\frac{L}{4}\Big),\frac{1}{h}F\Big(u = -\frac{L}{4}\Big),0\Biggr).
\label{ConstsAB}
\end{eqnarray} 

To study the total power emitted per mode $P_n$ of the decayed string loop, we make use use the main pieces of the polarization tensors, which can be integrated to find the total power emitted per mode \cite{gr-qc/0009091}:
\begin{equation}
\frac{dP_n}{d\Omega}=\frac{GM^2}{2\pi}\omega_n^2\Big\{ \langle{|{A_+^{(n)}}|^2}\rangle + \langle{|{A_\times^{(n)}}|^2}\rangle\Big\}.
\label{Psteradian}
\end{equation}
Eq. (\ref{Psteradian}) is our primary means of computing $P_n$. To verify our results, this quantity was also computed numerically using the stress-energy tensor for a string in Ref. \cite{1987PhRvD..36.2229G}.

	Consider first the special case of the square loop.  For this case, the loop retains its shape at arbitrarily late
	times, so we can evolve forward to any value of $N$.  Furthermore,
	by finding the point at which the ranges of $\tilde{u}$ and $\tilde{v}$ in Eq. (\ref{scaling}) are minimized, we note that the string decays totally away after $N$ oscillations, with $N$ given by
\begin{equation}
NG\mu=\frac{-1}{\frac{\ln w^2}{h^2}+\frac{\ln h^2}{w^2}},
\label{decayEq}
\end{equation}
corresponding to $N G\mu = 1/\ln(16) = 0.36$. 
	
\begin{figure}[H]
		\centering
		\includegraphics[width=0.8\textwidth, trim = 2cm 8cm 2cm 8cm, clip]{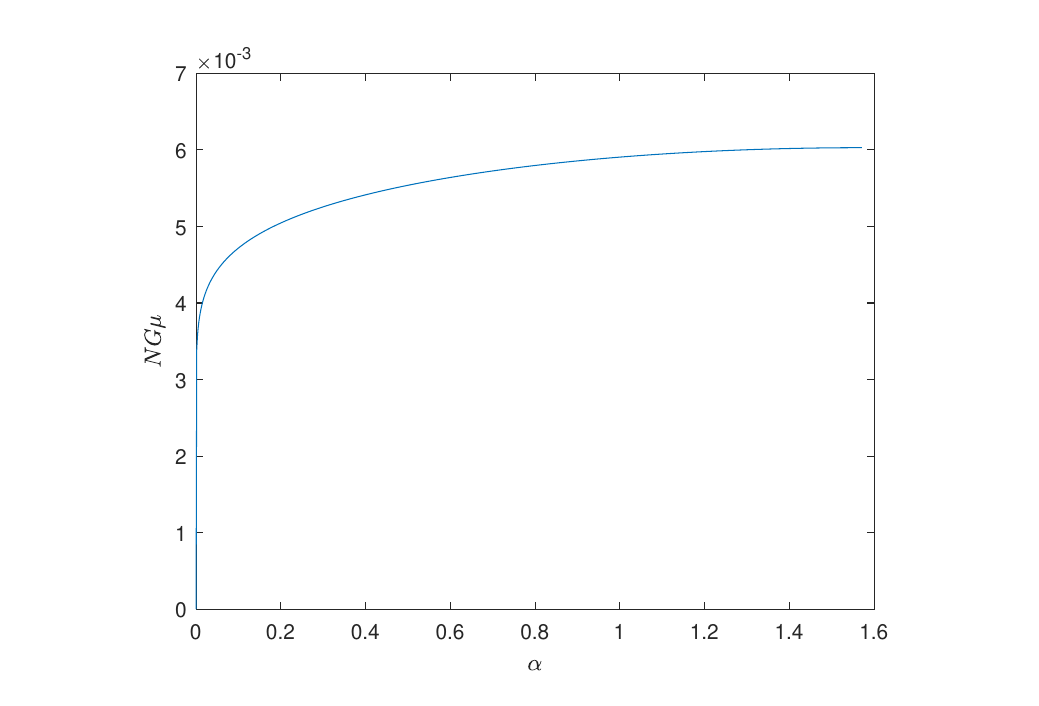}
		\caption{The maximum value of $NG\mu$ for which $\left|\Delta B'^0\right| < 0.1$
		so that Eq. (\ref{NGmurequirement}) is satisfied.}
		\label{fig:maxNGmu}
	\end{figure}

	In what follows, we take $\delta=10^{-5}$. The results for $P_n$ for the square loop ($\alpha = \pi/2)$ are exactly the same as in the undecayed case until total loop decay at $N G\mu = 0.36$.	For all of the other GV loops, $\alpha < \pi/2$, so gravitational radiation will smooth the kinks, changing the shape of the worldsheet for which the backreaction has been calculated. Therefore, the backreaction will need to be recalculated at some point (after some $N$ oscillations) due to the new shape. However, before this occurs, the 
	condition for the validity
	of the the first-order approximation (Eq. \ref{NGmurequirement}) will be violated.
	Taking (somewhat arbitrarily) $\left|\Delta B'^0\right| < 0.1$ to satisfy Eq. (\ref{NGmurequirement}), we see that the
	first-order approximation breaks down
	for $\max{\tilde{u}}=1-\delta$ and $\alpha=\pi/6$ around $NG\mu = 6
	\times 10^{-3}$, as can be seen in Fig. \ref{fig:maxNGmu}.
	
Therefore, we cut off our calculations at $NG\mu = 7 \times 10^{-3}$, which is a fiducial value used in Ref.
\cite{1609.01685}. The plots for the decaying loops are then shown for $NG\mu  < 7 \times 10^{-3}$ for a variety of opening
angles $\alpha$ in Figs. 4-7.
	
	\begin{figure}[H]
		\centering
		\includegraphics[width=0.8\textwidth, trim = 2cm 8cm 2cm 8cm, clip]{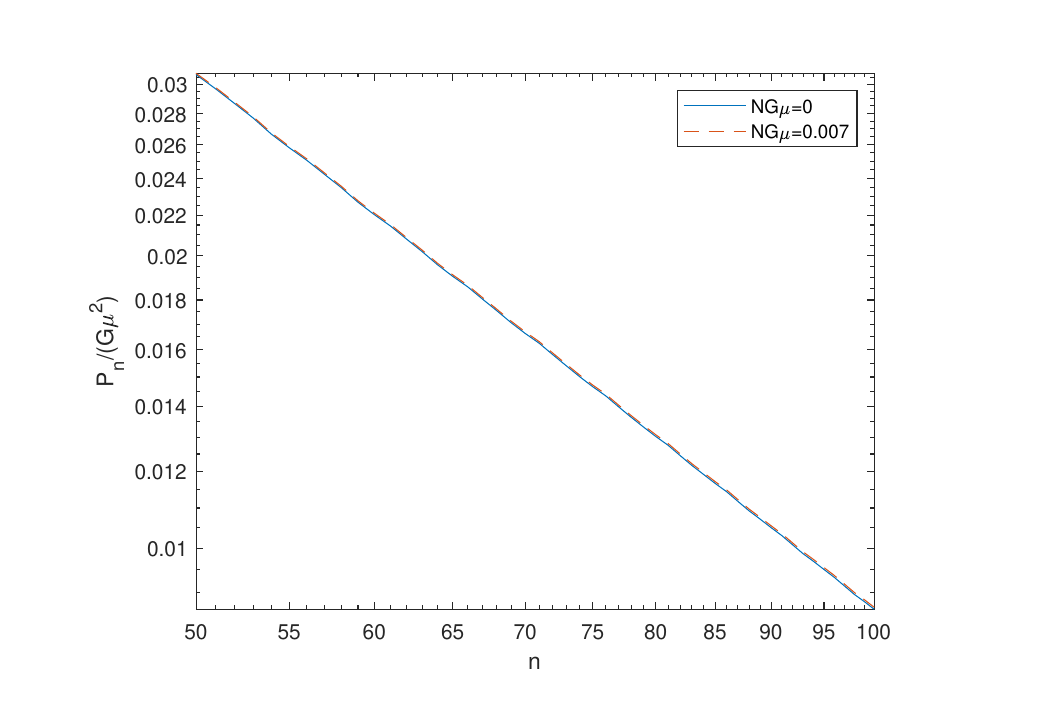}
		\caption{The power spectrum $P_n$ as a function of wavenumber $n$ for decaying GV
			loops with the opening angle $\alpha=5\pi/12$.}
		\label{fig:decayed_power_5pi12}
	\end{figure}
	\begin{figure}[H]
	\centering
	\includegraphics[width=0.8\textwidth, trim = 2cm 8cm 2cm 8cm, clip]{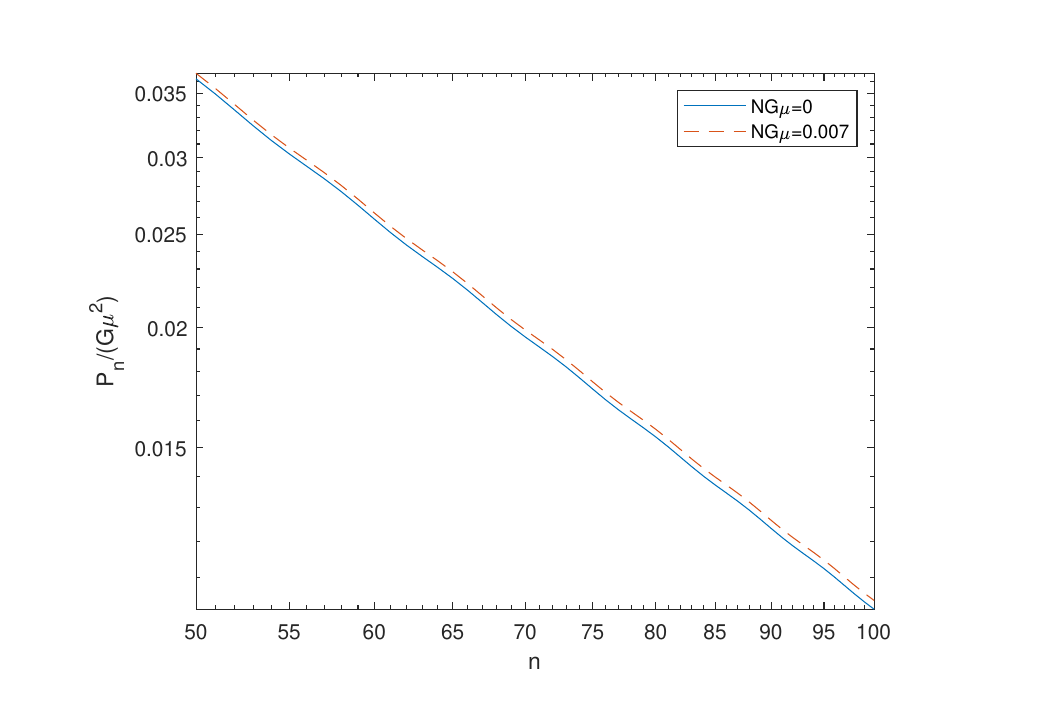}
	\caption{The power spectrum $P_n$ as a function of wavenumber $n$ for decaying GV
		loops with the opening angle $\alpha=\pi/3$.}
	\label{fig:decayed_power_pi3}
	\end{figure}
	\begin{figure}[H]
	\centering
	\includegraphics[width=0.8\textwidth, trim = 2cm 8cm 2cm 8cm, clip]{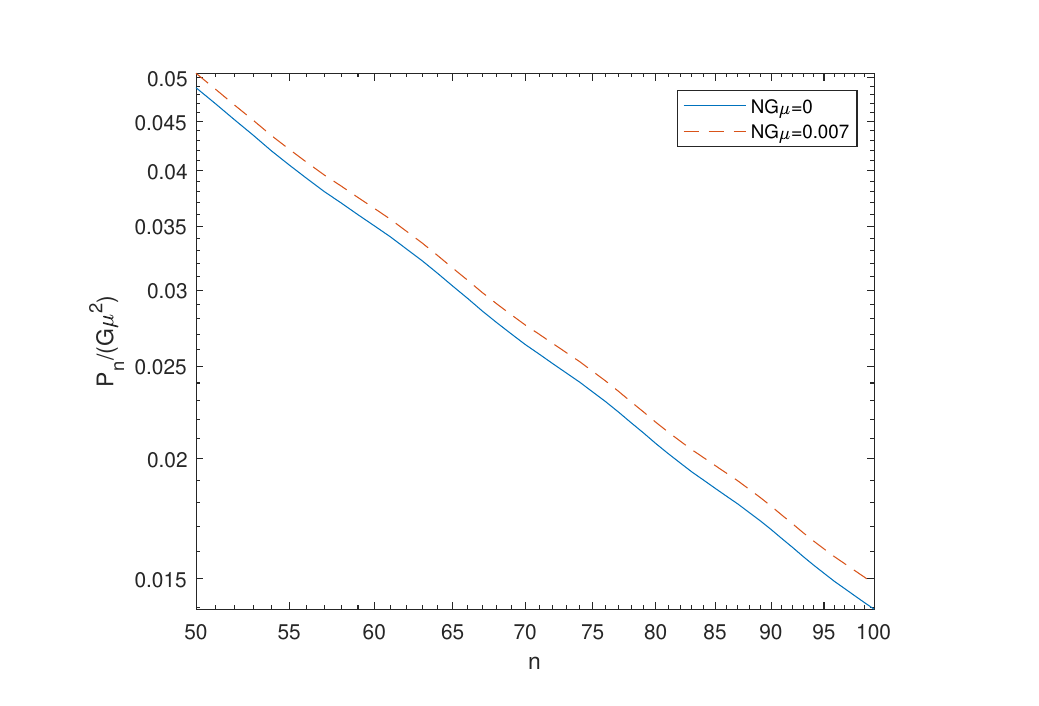}
	\caption{The power spectrum $P_n$ as a function of wavenumber $n$ for decaying GV
		loops with the opening angle $\alpha=\pi/4$.}
	\label{fig:decayed_power_pi4}
	\end{figure}
	\begin{figure}[H]
	\centering
	\includegraphics[width=0.8\textwidth, trim = 2cm 8cm 2cm 8cm, clip]{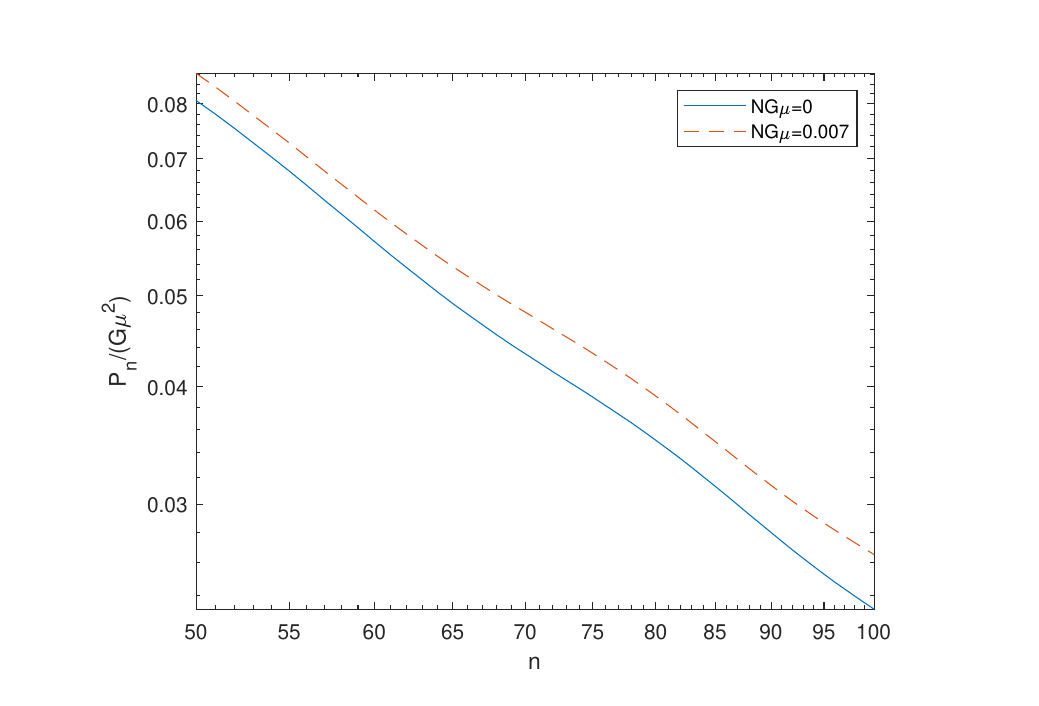}
	\caption{The power spectrum $P_n$ as a function of wavenumber $n$ for decaying GV
		loops with the opening angle $\alpha=\pi/6$.}
	\label{fig:decayed_power_pi6}
	\end{figure}
	
Note that for sufficiently square loops, the loop will decay before it significantly changes shape.  This occurs for $\alpha > 3 \pi/8$ \cite{1609.01685}.  For loops with a smaller value of $\alpha$, the effect of decay is to increase $P_n$ at large $n$ (shown in the figures) relative
to $P_n$ at smaller $n$ (below the lower cutoff for $n$ displayed in the figures).  This is apparent even over the relatively small values of $NG \mu$ over which the approximation
of Ref. \cite{1609.01685} is valid.

\section{Discussion}

Our most important result is that the initial power spectrum $P_n$ for GV loops scales as $n^{-2} \ln n$, rather than the expected $n^{-2}$.  To the extent that this result applies to the evolved distribution of cosmic string loops,
it would result in a bluer spectrum of relic stochastic gravitational radiation.  Furthermore, the lowest-order effect of loop
decays is to increase $P_n$  at large $n$ relative to its value at small $n$, making the the spectrum even more blue.
However, there are several major caveats to this conclusion.

The most significant caveat is the extent to which our results for GV loops can be extended to the evolved
loop distribution at all; i.e., to what extent do the loops produced in numerical
simulations correspond to cosmic strings
that are nearly planar with four kinks?
Consider first the numerical simulations of loop fragmentation derived by Copi and Vachaspati \cite{CV}.  In these simulations,
roughly 80\% of the loops contained between 3 and 5 kinks, with roughly 1/3 having
exactly 4 kinks as in the case of GV loops.  By analyzing the eigenvalues of the moment of inertia
tensors for the loops, Copi and Vachaspati showed that they tended to evolve toward nearly planar
configurations, which is also characteristic of GV loops. 
Blanco-Pillado et al. \cite{BOS1} found a loop distribution
that was more complex.  Prior to significant loop decay, the loops were cuspless, consisting of nearly
straight segments joined at kinks.  These
loops, like those in Ref. \cite{CV}, were largely planar.  However, they did not mimic the GV loops,
generally having more than 4 kinks.
While the number of kinks due to multiple fragmentations must necessarily approach 4 as discussed in Sec. II, this result
assumes multiple generations of loop fragmentation; this is likely the main difference between the simulations
in Ref. \cite{CV} and those in Ref. \cite{BOS1}.  Thus, it is unclear to what extent the results derived here are directly
applicable to the true loop distribution.

Further, it may be that our logarithmic correction is a result only of the high symmetry of the
GV loop trajectories.  It would be interesting, for example, to determine if this result carries over to more generic planar loops
consisting of straight segments, a
possibility we are currently investigating.
Of course, a logarithmic factor represents a small correction to a power law, with a correspondingly small
effect on the final predicted stochastic gravitational radiation background.  Nevertheless, this is an unexpected result,
making this a topic worthy of further investigation.

\acknowledgments
We thank K.D. Olum, J.M. Wachter, and T. Vachaspati for helpful discussions.

	\appendix
	
\section{Analytic power spectrum for square Garfinkle-Vachaspati loops}

Here we integrate Eq. (\ref{Pn}) from Ref. \cite{1987PhRvD..36.2229G} analytically
to derive an expression for $P_n$ valid in the limit of large $n$.  We restrict our discussion to square loops and consider
only even values of $n$.  Beginning with Eq. (\ref{Pn}) for even $n$, we have
\begin{equation}
P_n = 
\frac{32 G\mu^2}{\pi^3 n^2} \int d\Omega \frac{[1- \cos(n \pi e_1)][1- \cos(n \pi
e_2)]}{(1-e_1^2)(1-e_2^2)}, 
\end{equation}
which we decompose as
\begin{equation}
\label{decompose}
{P_n} = \frac{32 G\mu^2}{\pi^3 n^2}(I_1 + I_2 + I_3),
\end{equation}
with
\begin{eqnarray}
I_1 &=& \int d\Omega\frac{1- \cos(n \pi e_1)}{(1-e_1^2)(1-e_2^2)},\\
I_2 &=& \int d\Omega\frac{1- \cos(n \pi e_2)}{(1-e_1^2)(1-e_2^2)},\\
I_3 &=& \int d\Omega\frac {\cos(n \pi e_1)\cos(n \pi e_2) - 1}{(1-e_1^2)(1-e_2^2)}.
\end{eqnarray}
This decomposition has been chosen so that each of the three integrals is convergent.
By symmetry, $I_1 = I_2$, so we need only evaluate $I_1$ and $I_3$.

Recall that $e_1 = \hat r \cdot \hat{\bf a}$ and $e_2 = \hat r \cdot \hat{\bf b}$ where,
for a square loop, $\hata$ and $\hatb$ are perpendicular unit vectors.
To evaluate the first integral, we choose spherical coordinates such that $\hata$ lies on the $z$ axis and $\hatb$ lies
on the $y$ axis, so that $e_1 = \cos(\theta)$ and $e_2 = \sin(\theta)\sin(\phi)$.
Making the standard substitution $z = \cos \theta $, and taking, by symmetry, twice the integral
over the upper hemisphere, $I_1$ becomes
\begin{equation}
I_1 = 2 \int_{z = 0}^{1} \int_{\phi = 0}^{2\pi}\frac{1- \cos(n \pi z)}
{(1-z^2)[1 - (1-z^2)\sin^2 (\phi)]} dz ~d\phi.
\end{equation}
Taking the integral over $\phi$, we get
\begin{equation}
I_1 = 4 \pi \int_{z = 0}^{1} \frac{1- \cos(n \pi z)}{z-z^3}.
\end{equation}
We can integrate this (using $\sin(n\pi) = 0$ and $\cos(n\pi) = 1$ for even $n$) to obtain
\begin{equation}
\label{I1intermediate}
I_1 = 2 \pi\left[\Ci(n\pi (1+z)) - 2 \Ci(n\pi z) + \Ci(n\pi (1-z)) + 2\ln(z) - \ln(1+z)-\ln(1-z)\right]\biggr|_0^1,
\end{equation}
where
$\Ci(x) \equiv -\int_x^\infty (\cos t/t) dt$ is the cosine integral function.

To make further progress, we assume that $n \gg 1$.  This allows us to use two
properties of the cosine integral, namely $\Ci(x) =  \gamma + \ln(x) + O(x^2)$
as $x \rightarrow 0$, and $\Ci(x) \rightarrow 0$ as $x \rightarrow \infty$.
Here $\gamma$ is the Euler-Mascheroni constant ($\gamma \approx 0.577$).  Then in the limit of large $n$, Eq.
(\ref{I1intermediate})
reduces to
\begin{equation}
\label{I1final}
I_1 = 2 \pi [3 \gamma + 3 \ln(\pi) - \ln(2) + 3 \ln(n)].
\end{equation}
Since $I_2 = I_1$, we now need only evaluate $I_3$.

We first rewrite $I_3$ as
\begin{equation}
\label{I3firstmod}
I_3 = \int d\Omega\frac {\cos[n \pi (e_1 + e_2)]   - 1}{2(1-e_1^2)(1-e_2^2)}
+ \int d\Omega\frac{\cos[n \pi (e_1 - e_2)] - 1}{2(1-e_1^2)(1-e_2^2)}.
\end{equation}
For the first integral in Eq.(\ref{I3firstmod}) we choose a spherical coordinate system for which
$\hata = (1/\sqrt 2)(\hat z + \hat y)$
and $\hatb = (1/\sqrt 2)(\hat z - \hat y)$, while for the second integral,
we take a coordinate system for which
$\hata = (1/\sqrt 2)(\hat z + \hat y)$
and $\hatb = (1/\sqrt 2)(-\hat z + \hat y)$.
Then Eq. (\ref{I3firstmod}) becomes
\begin{eqnarray}
\label{I3secondmod}
I_3 &=& \int d\Omega\frac {\cos\left[\sqrt 2 n \pi  (\cos \theta)\right]   - 1}{2\left[1-\frac{1}{2}(\cos\theta + \sin\theta \sin
\phi)^2\right]
\left[1-\frac{1}{2}(\cos \theta - \sin \theta \sin \phi)^2\right]} \nonumber \\
&+& \int d\Omega\frac{\cos[\sqrt 2n \pi  (\cos \theta)] - 1}{2\left[1-\frac{1}{2}(\cos\theta + \sin \theta \sin \phi)^2\right]
\left[1- \frac{1}{2}(-\cos\theta + \sin \theta \sin \phi)^2\right]}.
\end{eqnarray}
It is clear that the two integrals are equal, so we can write (taking $z = \cos \theta $ and $s = \sin \phi$ and
using the symmetry of the integral),
\begin{equation}
I_3 =  4 \int_{z = -1}^{1} \int_{s = 0}^{1} \frac {\cos\left(\sqrt 2 n \pi  z\right)  - 1}{\left[1-\frac{1}{2}(z + \sqrt{1-z^2} s )^2\right]
\left [1-\frac{1}{2}(z - \sqrt{1-z^2} s)^2\right]}\frac{ds}{\sqrt{1-s^2}} dz. 
\end{equation}
The integral over $s$ yields
\begin{equation}
I_3 = 4\pi \int_{v=0}^{\sqrt{2}}\frac{\cos(n \pi v)-1}{v}\left[\frac{1}{(2-v)|1-v|} - \frac{1}{(v+2)(v+1)}\right]dv,
\end{equation}
where we have used symmetry to take twice the value of the integral over the upper hemisphere, and we have
made the substitution $z = v/\sqrt 2$ to simplify our expression.
Splitting the integral into the ranges $0 < v < 1$ and $1 < v < \sqrt 2$ gives
\begin{eqnarray}
I_3 &=& 4\pi \int_{v=0}^{1}\frac{\cos(n \pi v)-1}{v}\left[\frac{1}{(2-v)(1-v)} - \frac{1}{(v+2)(v+1)}\right]dv \nonumber\\
&+& 4\pi \int_{v=1}^{\sqrt{2}}\frac{\cos(n \pi v)-1}{v}\left[\frac{1}{(2-v)(v-1)} - \frac{1}{(v+2)(v+1)}\right]dv,
\end{eqnarray}
and using $\sin(2 \pi n) = \sin (\pi n) = 0$
and $\cos(2 \pi n) = \cos(\pi n) = 1$ for even $n$, we get
\begin{eqnarray}
I_3 &=& 2 \pi [\Ci(n \pi (2-v)) - 2  \Ci(n\pi(1-v))
+ 2 \Ci(n\pi(v+1))
- \Ci(n\pi(v+2)) \nonumber\\
&+&2\ln(1-v) - \ln(2-v)-2\ln(v+1)+\ln(v+2) \biggr|_0^1 \nonumber \\
&+& 2 \pi [-2\Ci(n\pi v) -\Ci(n \pi (2-v)) + 2\Ci(n\pi(v-1)) + 2 \Ci(n\pi(v+1)) -\Ci(n\pi(v+2)) \nonumber \\
&-&2\ln(v-1) + \ln(2-v)-2\ln(v+1)+\ln(v+2) + 2 \ln v]
\biggr|_1^{\sqrt 2}.
\end{eqnarray}
Taking the limit of large $n$ and using the asymptotic behavior of the cosine integral
function discussed above, we obtain
\begin{equation}
\label{I3final}
I_3 = 2\pi[-4 \gamma - 4 \ln \pi - 4 \ln(n) + 2 \ln(2)].
\end{equation}
Now we have $P_n = ({32 G\mu^2}/{\pi^3 n^2})(2I_1 + I_3)$, and inserting the values of $I_1$ and $I_3$
from Eqs. (\ref{I1final}) and (\ref{I3final}) gives us the expression for $P_n$ for square loops in
the large-$n$ limit for even $n$:
\begin{equation}
\label{finalPn}
P_n = \left[ \frac{128 G\mu^2}{\pi^2 n^2}\right ] \left[\gamma + \ln \pi
+ \ln(n)\right].
\end{equation}

It is clear from this discussion that we could derive an exact expression (valid
for all $n$, not just the large-$n$ limit) in terms of sums of cosine integral
functions, but the resulting expression is quite cumbersome and does not provide
much insight into the behavior of $P_n$.  For odd $n$, a decomposition as in Eq. (\ref{decompose})
is not possible, because the integrand in each of the separate terms becomes singular at $e_1 = 1$, $e_2 = 0$
and $e_1 = 0$, $e_2 = 1$.  However, our numerical results suggest that Eq. (\ref{finalPn}) is valid
for odd $n$ as well.

\end{document}